# The Power of Internet of Things (IoT): Connecting the Dots with Cloud, Edge, and Fog Computing

Shams Forruque Ahmed[1,*], Shanjana Shuravi[1], Shaila Afrin[1], Sabiha Jannat Rafa[1], Mahfara Hoque[1], Amir H. Gandomi[2,3,*]

[1] Science and Math Program, Asian University for Women, Chattogram 4000, Bangladesh
[2] Faculty of Engineering & Information Technology, University of Technology Sydney, NSW, 2007, Australia
[3] University Research and Innovation Center (EKIK), Óbuda University, 1034 Budapest, Hungary

*Corresponding authors: shams.ahmed@auw.edu.bd, shams.f.ahmed@gmail.com (Shams Forruque Ahmed); gandomi@uts.edu.au (Amir H. Gandomi)

## Abstract

The Internet of Things (IoT) is regarded as an improved communication system that has revolutionized traditional lifestyles. To function successfully, IoT requires a combination of cloud, fog, and edge computing architectures. Few studies have addressed cloud, fog, and edge computing simultaneously, comparing them and their issues, although several studies have looked into ways of integrating IoT with either one or two computing systems. Thus, this review provides a thorough understanding of IoT integration with these three computing architectures, as well as their respective applications and limitations. It also highlights the advantages, unresolved issues, future opportunities and directions of IoT integration with the computing systems to advance the IoT. IoT can use the Cloud's almost limitless resources to overcome technology restrictions, such as data processing, storage, and transmission. While edge computing can outperform cloud computing in many circumstances, IoT and edge computing become increasingly integrated as IoT devices increase. Cloud computing also poses a few issues, including managing time-sensitive IoT applications like video gaming, simulation, and streaming, which can be addressed by fog computing integrated with IoT. Due to the proximity of fog computing resources to the edge, data transfers and communication delays to the cloud can be reduced as a result of combining the two. The integration of IoT with cloud, fog, and edge computing will create new business prototypes and opportunities. Since IoT has the potential to greatly enhance connectivity infrastructure as an inevitable component of the future internet, further study is needed before it can be fully integrated.

*Keywords:* Internet of things; IoT; Edge computing; Cloud computing; Fog computing

## 1. Introduction

The Internet of Things (IoT) is becoming a milestone in the field of Artificial Intelligence (AI) and a step towards a revolution in the current world [1]. "Things" refers to physical devices in this context; hence, this network is composed of sensors and processors that supports data communication between devices [2]. This system has two kinds of nodes: physical nodes and virtual nodes. Sensors, actuators, transit nodes, and other wearable or embedded devices are examples of physical nodes in the IoT, whereas virtual nodes include virtual machines or networks utilized by a group of wireless networks [3]. An IoT architecture consists of sensors, protocols, actuators, cloud services, and layers, among other components. The three layers of this architecture are essential for evaluating data, providing insights, identifying industrial risks, and delivering fast solutions for any of the linked devices. These IoT tasks are performed by combining three architectures: cloud, fog, and edge computing. IoT device analytics and monitoring are made possible by the cloud computing architecture, which provides essential application-specific services across a wide range of functional domains for IoT device analytics and monitoring [4]. In contrast, fog and edge computing architecture are crucial for nodes to execute real-time data processing and data computation at the network's edge [5]. The cloud computing architecture, also known as utility computing, provides flexible network access for scalable, QoS-assured services on demand [6].

Although IoT and cloud computing are developed in different ways, their combination introduces the concept known as the cloud of things. Based on fundamental components, such as cloud-to-device interface, authentication, database administration, and cloud-to-user interface, the design and implementation of an IoT-Cloud platform were analyzed by Bhawiyuga et al. [7]. To perform the analysis, all modules were installed on a virtual private server (VPS) with the following specifications: 30 GB SSD drive, 1 GB RAM, 1.6 GHz single-core CPU, public IP address, Apache 2.4.7 web server, and Django Framework for constructing RESTful HTTP services. The suggested method was able to provide communication, security, and storage features in functional testing. The performance results demonstrated an increase in the number of concurrent connectivity options, which influences the delay in the IoT device receipt from the cloud system. Device to cloud interfaces often serves as a data transmission endpoint between cloud services and IoT devices [8]. As such, combining cloud and IoT services can maximize resource utilization in some cases. While cloud providers enable data transfer through the internet, making data navigation easier, fog computing permits IoT devices to process, decide, and act to transmit relevant data to the cloud. In many circumstances, fog and edge computing architectures appear functionality identical and are both cloud-based data transmission boosters, yet there are differences [9]. Fog computing works as a gateway between the edge and the cloud, whereas edge computing focuses on data processing. Fog computing also facilitates the optimization of services and the development of improved user interfaces, such as faster responses for time-sensitive applications.

Although cloud computing is a well-known method for analyzing and providing results for massive volumes of data in IoT, it presents some drawbacks that can be alleviated by fog and edge architectures. Over the past few years, numerous studies have determined that IoT systems that rely solely on a subset of a single architecture, such as cloud, fog, and edge computing, are unlikely to respond in real time or gain useful insights from data. According to Farshad Firouzi [10], edge and fog computing were developed to facilitate the interaction between the cloud and IoT, which distributes data processing resources to the data sources' edges in addition to the cloud. The authors reported that the layered and collaborative edge–fog–cloud topology offers significant advantages since it allows the dispersion of intelligence and computation. The combination of cloud, fog, and edge architecture can therefore be viewed as aiding IoT by enhancing data computation allocation and minimizing network traffic, resulting in improved operational efficiency.

Despite the usage of the three computing architectures, some areas of computing networks still suffer from the following issues [11]: faults in 5G network infrastructure, data storage inaccuracy in industrial IoT, resource allocation, optimization error, increased energy consumption, and complexities in business and service models, among others. In this age of fast communication systems, IoT technology is necessary for various fields, such as e-commerce, industrial infrastructure, data security, the development of new business models, and others. Despite the advancement of many IoT applications in the past few years, the technology is still in its early stage, which may lead to the issues mentioned earlier. As illustrated in Table 1, recent review studies [12][13][14][15][11][16][8][17] on IoT integration identified numerous challenges with this communication technology, but did not provide insights into how cloud, fog, and edge might be used simultaneously to overcome those challenges. Most of these works only focused on the integration of IoT with a single computing architecture and/or its applications. This article thus provides a comprehensive overview of IoT integration with the three computing architectures and their respective applications and issues. In order to support IoT advancement, this review aims to contribute an in-depth understanding of IoT integration with the cloud, fog, and edge computing as well as to highlight unresolved issues.

Table 1. A comparison of recent relevant reviews (2018-2022) with the current one, which integrates IoT with cloud, edge and fog computing

| Review study | IoT architectures | IoT integration with cloud computing | IoT integration with edge computing | IoT integration with fog computing | Applications of IoT computing integration | Advantages and challenges of IoT integration with Cloud, Edge and Fog computing |
|---|---|---|---|---|---|---|
| This study | √ | √ | √ | √ | √ | √ |
| Ali et al. [8] | √ | × | √ | √ | × | × |
| Bittencourt et al. [11] | √ | √ | × | √ | √ | × |
| Atlam et al. [12] | × | × | × | √ | √ | √ |
| Bellavista et al. [13] | × | × | × | √ | √ | × |
| Jahantigh et al. [14] | √ | √ | × | × | √ | × |
| Dizdarević et al. [15] | √ | × | √ | √ | × | √ |
| Kong et al. [16] | √ | × | √ | × | √ | √ |
| Laroui et al. [17] | √ | × | √ | √ | √ | × |

√: sufficient information available; ×: sufficient information unavailable

## 2. Internet of things and their progress

The IoT is a network of interconnected computing devices that allows computers, mechanical and digital equipment, items, animals, and even people to exchange data with one another and with other devices through the internet without the need for human intervention [18]. The internet and smart devices are the primary tools for constructing and operating the IoT networking infrastructure. Both cloud computing and mobile technologies enable the sharing and collection of big data and analytics with minimal human intervention. This digital system can be used to control, monitor, and record the interactions between connected devices. Through the use of quantum and nanotechnology, IoT has established a framework for storage, sensing, and processing speed that was previously unimaginable. IoT is consequently considered as an essential component of 21st century technology breakthroughs that brings innovative solutions to a variety of global challenges [19]. IoT has already demonstrated its potential through its widespread adoption in human services. It is currently utilized in a range of industries, including the health care industry, environmental business, and consumer-oriented enterprises, and may offer a large array of functionalities in the future. This will generate several employment opportunities and higher wages. The continuous development and expansion of IoT are introducing plenty of new possibilities for research, development, and employment [20]. Since the IoT facilitates both the interconnection and intercommunication of large, ubiquitous things, it is causing an unanticipated output of massive and diversified amounts of data known as data explosions [21]. Security is another issue with IoT as an unlimited number of unsecured devices can connect to it, and thus, improving IoT is required to secure the algorithm used for encryption [22].

IoT has made a significant contribution to the realm of technology, adding a new dimension that allows us to make connections worldwide. IoT-based devices and equipment are largely used today because they are convenient, pertinent, relatively inexpensive to manufacture, and consequently, generally available within a reasonable price range. IoT has not only demonstrated its potential, but it has also extended and revealed its significance and worth in the revolutionary development of the trading and stock exchange markets [23]. However, the security and privacy of information and data appear to be a big concern and challenging issue to be addressed in this regard [24]. In order to illustrate and highlight the potential utility and applicability of IoT infrastructures in various domains [25], comprehensive research studies have been undertaken [26][27][28]. Some of the most prominent IoT innovations include smart home systems (SHS) consisting of internet-based gadgets, smart health sensing systems (SHSS) that are made of small intelligent equipment to support human health, home automation systems, as well as reliable energy management systems. Kim et al. [27] designed and presented IoT health care service guidelines based on the user's experience with the service providers. In order to verify the services, an analysis was performed to examine the proposed parameters like risk sensitivity and trust. The study identified some prominent key factors that influenced consumers' confirmation of those services and reported that people in South Korea prioritize reliable and secure healthcare services, which mainly concentrated on lifestyle diseases. It also provided a straightforward and comprehensive direction for IoT healthcare service companies to ensure reliability. However, the collected data is mostly a hypothetical description of services rather than an actual commercial healthcare service used by consumers. The system's performance has also not been evaluated. Another study presented by Ding et al. [29] highlighted the IoT-based remote medical monitoring system on mobile devices. With the help of artificial cognitive nodes, the authors proposed a framework of interpersonal interaction and psychological parameters that can be used to develop effective alarm systems for any kind of emergency by preserving essential emergency information in the hospital database. The framework continued to improve the precision, accuracy, and rapidity of measuring the psychological parameters of low-power consumption devices being used in this regard. However, the expenses were not taken into account in the study.

A prototype was developed by Jimenez et al. [30] for an IoT healthcare monitoring system that can be used for longer periods of time. The system was developed by utilizing low-cost sensors and current IoT platforms, which are commonly found in households worldwide. During the system execution period, the suggested ad-hoc monitoring system can examine a patient's biological parameters [23], while several sensors of the system can measure meteorological data, such as temperature and humidity. Gadgets like smartphones are mostly used in the patient's environment as gateways to acquire relevant data from sensors. In order to evaluate the prototype's scalability and performance, the recommended healthcare monitoring system was used to collect business needs. This prototype was constructed using low-cost sensors and energy-efficient gadgets. However, it provided the bare minimum of service in accordance with the business requirements. In addition, IoT has had a massive impact on environmental sustainability by introducing IoT-based monitoring systems in agriculture, such as SmartFarm, carbon dioxide and climate change monitoring, as well as regional waste and disposal management [31]. Muangprathub et al. [32] proposed an IoT-based wireless sensor network by developing and deploying a monitoring system among node sensors in farms, which could be utilized for watering crops and optimizing agriculture. This framework was basically designed to stimulate and enhance the production of different crops. This hybrid framework of mobile and web applications, using LINE API for mobile user interfaces, was connected to IoT nodes installed in different areas. Data transmission among the nodes can be handled through a smartphone and web-based applications. A prominent automated technique of data mining was implemented, and the collected data were analyzed to predict the appropriate temperature, humidity, and soil quality. The results demonstrated the efficacy of implementing this network, which will eventually be suited for establishing a smart farm.

Smart cities are now being developed with the help of IoT applications, which enable remote monitoring, management, control of devices, and to generate new approaches by extracting from a huge amount of actual data [33]. Two of the most significant aspects of a smart city are extensive IT integration and application [34]. The installation of RFID, IR, scanning and GPS sensors can lead to the development of a smart city that achieves the objectives of communication, information transfer, monitoring, and management [35]. Lin et al. [36] reported the implementation of various IoT applications in various locations in Taiwan. The applications were established and developed with the aid of the IoT device management platform "IoTtalk." Four applications were location-based, including: i) a dog tracking program that tracks a dog's movements; ii) an emergency software application that offers service for any type of emergency; iii) the use of multiple PM2.5 sensors; and iv) the use of a robot comprised of sensors to collect building information, such as indoor environment quality and temperature. Instead of detecting the sensors, this study provided an appropriate way for recognizing the locations of sensors, which includes updating the IoTtalk GUI setting option for different places. The research developed a technique for identifying IoT devices while optimizing the balance between energy usage and position tracking. Nevertheless, it did not demonstrate any particular sensor-related solution.

IoT-Sim, based on Cloudsim, was constructed by Zeng et al. [37] to enable the simulation of IoT-based data processing as well as MapReduce through cloud computing. Based on the expense of maintaining large-scale data centers using the Big Data processing framework, this study indicated that inspecting and evaluating applications of IoT in an original cloud computing environment is quite challenging. The offered simulator provides a suitable environment for studying and assessing IoT applications on top of Cloudsim. Due to the limits of dealing with poor latency and the actual requirements of such applications, the proposed simulator is unable to process streams in MapReduce. Zia et al. [38] developed an application-based forensic exploratory digital approach and evaluated it in three typical IoT application scenarios, including smart home, wearable, and smart city. In this study, digital forensics in IoT was recommended for evaluating industry practices to execute the developed model. Using the IoT-based forensic application, the proposed model is believed to be useful in any forensics investigation to evaluate and generate forensics reports. The presented methodology has the advantage of being able to analyze a wide range of forensic evidence in several IoT systems. However, IoT security protocols were not considered, even though such protocols would be beneficial to make the approach more practical and secure.

## 3. Existing IoT architectures

Mastering any computer network requires familiarity with its underlying technology. Since IoT devices are resource-constrained, their offered services will impose precise criteria, security being the most crucial. Consequently, understanding and conforming to these standards require illuminating the fundamental architecture of IoT and its affiliated aspects. IoT architecture works as a procedure of data circulation from sensors linked to "things" to a corporate data center or the cloud through a network for processing, analyzing, and storing. A "thing" in the IoT could be a structure, machine, building, or even a human [39]. It is considered a virtual, physical, or hybrid system devised of several functional physical objects, actuators, sensors, cloud services, customized IoT protocols, users, developers, communication layers, and an enterprise layer. Specialized architects function as a vital aspect of the IoT foundation, providing a systemized method for diverse components that leads to solutions to correlated problems [40]. Thus far, some of the available IoT architectures are three-layer, four-layer, five-layer, cloud-based, Service Oriented Architecture (SOA), fog-based, and software-defined network (SDN)-based [39,41].

The majority of IoT systems are cloud-based and freely accessible. Amazon Web Services (AWS) leads the commercial cloud industry by offering a wide range of data processing services (AWSLambda, Amazon S3, Amazon SQS, Amazon Kinesis, Amazon DynamoDB, Amazon SNS). Another example of an IoT platform is Microsoft Azure IoT Hub, which has a reference architecture contrived of core platform services and software modules. It enables three prime areas of an IoT solution (device connectivity, analytics, data processing and management; data presentation, and business affinity). OpenMTC, FIWARE, and Site are a few other platforms mentioned in various studies [42,43].

### 3.1. Three-layer architectures

Only an efficient IoT architecture layer can ensure the best, most authentic, and most secure convergence of communication and information systems with the fastest speed. The three-layer architecture (Fig. 1) is one of the leading and elementary IoT architectures, which includes the following three layers: perception, network, and application. Although it is quite practical and realizable, it is unable to provide a feasible solution due to the complexity of IoT [44,45].

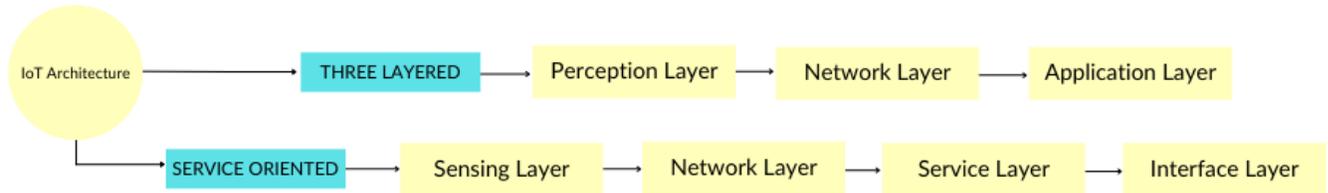

**Fig. 1.** Three-layered architecture and service-oriented architecture of IoT.

### 3.1.1. Perception layer

The perception layer is the bottom layer, also known as the control layer, that focuses on object recognition and data collection from users, requested services, and the environment. It is comprised of detection equipment like RFID, bar codes, distance sensors, and other physical objects [46]. The desired service will determine the most suitable monitoring gadget. This layer has sensing capabilities that allow it to capture and accumulate signals from the environment consisting of smart things [47]. After completing the data collection, this layer performs initial data processing and packaging, and also obtains control signals from the network layer to accomplish the required control executions via the executive gadgets. Some data sensors contained in this layer include Wireless Sensor Network (WSN), Reactive Sensor Network (RSN), Radio Frequency Identification (RFID), and actuators [48]. The perception layer aims to detect distinct objects and interact with the received information, such as humidity, location, vibration, amount of dust in the air, pH level, wind speed, vibration, etc. Then, this data is sent through the network layer to the central information processing system for safe interaction. A vast percentage of terminals accumulate the required information instantaneously and convey it to the user or form an opinion on things. Leakage of secret information, terminal virus, tampering, and copying are among the primary challenges in the perception terminals [49]. Since RFIDs and sensors configure the majority of the IoT perception layer, their power consumption, storage capacity, and computational capacity are all constrained, rendering them vulnerable to numerous attacks and threats [50]. Perhaps, using encryption, confidentiality, authentication, and access control can tackle these security concerns [49].

### *3.1.2. Network layer*

The network layer is also defined as the transmission layer that serves as a liaison appending the application and perception layers. It functions like a human brain and neural network, carrying data from physical objects and distributing them through sensors. This channeling approach can be wired or wireless and uses protocols like Wi-Fi, 3G, UMTS, Infrared, WiMAX, Satellite, Bluetooth, and ZigBee [51,52]. It is also in charge of clamping network devices, smart items, and networks to one another. As a result, it faces critical security vulnerabilities with data solidarity and authentication and is highly susceptible to attacks, such as:

- Denial of Service (DoS) attack prohibits authorized users from gaining access to devices and other network resources [53].
- Main-in-the-middle (MiTM) attack is a severe concern for online safety since it permits a hacker to grasp and reshape data instantaneously. It occurs when an invader surreptitiously expropriates or alters the transmission between a sender and recipient [54].
- In a Storage Attack, hackers attack both the cloud and storage device, which contain users' information, then modify the information erroneously [55]. Threats are more likely when data are reproduced and exposed to other data by multiple people.

### *3.1.3. Application layer*

The application layer is the top layer that analyzes the information from the network and perception layers, eventually resulting in the IoT application. This layer is accountable for giving application-specific help to the user. It defines various implementations to deploy IoT, for instance, in smart homes, smart health, smart eyewear, smart cities, and smart vehicles. It is the interface between IoT and diverse users (individuals or systems) with specialized needs to achieve various intelligent IoT applications (i.e., intelligent traffic, intelligent building, intelligent logistics, security monitoring, and vehicle navigation). Since these IoT applications entail information technology usage [45,56], the application layer ensures the validity, integrity, and confidentiality of information [50]. The key shortcoming of this architecture is that more activities will be allocated to a single layer, making it difficult to update single or several layers [57]. Some potential threats and challenges in the application layer include:

- Malicious Code Injection is an attack where specific hacking tactics exploit an end-user attack on a system, whereby the attacker can insert any form of mischievous code and steal information from the user.

- DoS attacks on the application layer have become more sophisticated in recent years. It provides a smokescreen for carrying out cyberattacks to infiltrate the defensive system and, thus, the user's data privacy while fooling the victim into thinking the attack is occurring elsewhere.

- Spear-Phishing Attack is an e-mail spamming attack in which the victim is tempted to open an email so that the attacker can gain access to their credentials.

- In a Sniffing Attack, a hacker can initiate an attack by installing sniffer software to obtain network information and corrupt the system [58].

### 3.2. Service-oriented architecture (SOA)

SOA is a paradigm of application frameworks, wherein software components use a communication protocol to serve via a network. This architecture provides an approach to model a vast software platform, utilizing web services, where computational components or sub-software are dispersed among multiple remote servers that serve other users. It unifies software components that are distributed, individually controlled, and deployed [59–61]. Despite the adaptability of SOA, scaling, integrating, and fortifying resilience persist as challenges in IoT systems. The deficiency of an intelligent connection-aware infrastructure is one of the crucial factors of this deficient integration in IoT systems [62]. Although IoT is becoming increasingly popular in numerous fields to make peoples' lives easier, integrating the physical and virtual worlds is a massive challenge [51]. There is no precise definition for SOA since it has been defined by several scholars from multiple viewpoints (i.e., technology, architecture, business). SOA is neither a technology, product, nor an expeditious way to deal with IT convolution [63] – it is viewed as a complicated system, well-defined simple item, or collection of subsystems. The hardware and software parts of IoT can be reused and enhanced effectively since those subsystems can be reused and managed independently [64]. The service provider, service requester, and service registry are the three core modules of SOA.

SOA is comprised of four layers (Sensing, Network, Service, and Interface (Fig. 1)) with distinct characteristics and facilitates device interoperability in several ways. The sensing layer is integrated to sense the status of all available objects. It attempts to cognize the information by applying the IP address and sending it to the network layer. It is also in charge of data processing, utilizing computational algorithms [57,64]. A Universal User Identifier (UUID) is a 128-bit integer used to identify a particular object or entity on the internet. Using this, one may uniquely identify objects and track the environment for a variety of applications. Size, energy consumption, resources, deployment, accessibility, retrievable, and cost should be considered when determining the sensor layer of an IoT [64].

The network layer is the base to assist and transfer data through wireless or wired connections [65]. The service layer is responsible for building and controlling connections based on user and client requests. All service-oriented operations (i.e., exchange and storage of information, data management, ontology databases, communication, and search engines) are implemented at this layer. The majority of the requirements are met by numerous standards created by individual organizations. A practical service layer is comprised of application programming interfaces (APIs), a set of required applications, and protocols that support obligatory services and applications. However, a universal service layer is necessary for the IoT architecture. The application layer is another appellation of the interface layer, which is accountable for data formatting and presentation. It is composed of interaction mechanisms with programs, users, and other applications with data about communication strategies [57,64]. Table 2 tabulates the key features and challenges/threats of the three-layer and service-oriented architectures.

SOA-based services are widely utilized in the development of large enterprises and, thus, play a central part in the IoT domain. Fresh and emerging resources are opening on the internet in the context of IoT. As a result, research on SOA-based fusion applications is significantly valuable. Several studies have discussed the use of SOA in different sectors, including healthcare, agriculture, activity recognition, decision-making, and the military [66–70]. Alsaryrah et al. [71] recognized SOA as the primary driver of IoT. They claimed that combining IoT services with SOA enhances the manufacture of value-added and intricated IoT applications by merging atomic services to provide adequate distinctive features. However, SOA is hindered by its high expense, high overhead, and complex service management [72].

**Table 2.** Key features and challenges/threats of three-layer architecture and service-oriented architecture.

| Architecture | Layers | Key Features | Challenges/Threats | Refs. |
|---|---|---|---|---|
| Three-layered Architecture | Perception layer | Senses physical parameters, gathers information and detects smart objects. | Constrained power consumption, storage capacity, and computational capacity make it vulnerable to threats and attacks. | [46,47,49] |
| | Network layer | Encapsulation sublayer composes the packets, and the routing layer transfers the packets to the destination. | DOS, MiTM, and storage attack | [51–54] |
| | Application layer | Works as an interface to link the communication between the network and IoT. | Malicious Code Injection, DOS, Sniffing Attack, and Spear-Phishing Attack | [50,57,58] |
| Service-oriented Architecture | Sensing layer | Detects information and transfers it to the network layer. | Requires high investment | [57,60,64,65,72] |
| | Network layer | Supports and transfers data through wireless or wired connections among other things. | Unreliable network channels and components can cause a loss of bits during data transfer. | |
| | Service layer | All service-oriented operations, such as exchange and storage repository of information, data management, ontology databases, communication, and search engines, are implemented at this layer. | There is no reliable universal service layer. | |
| | Interface layer | Responsible for data formatting and presentation. | As a bridge of IoT with user applications, it should fortify legal and trustworthy interaction with other applications or users. | |

## 4. Cloud, Edge and Fog computing architectures

IoT provides a means of universal computing in which devices with unique addressing systems can communicate and share data with one another. Uniquely, it allows things and people who use these devices to be linked under any condition, e.g., any place, moment, with anybody or anything, and communication and interaction with any system, network, path, service, or any type of communication [73,74]. The interaction among the computing architectures, such as cloud computing, fog computing, and edge computing in IoT, is depicted in Fig. 2. In brief, cloud computing architecture facilitates device-to-device and app-to-app communication in the IoT, while fog and edge computing architectures act as extensions of cloud networks, which are decentralized networks comprised of a collection of computers.

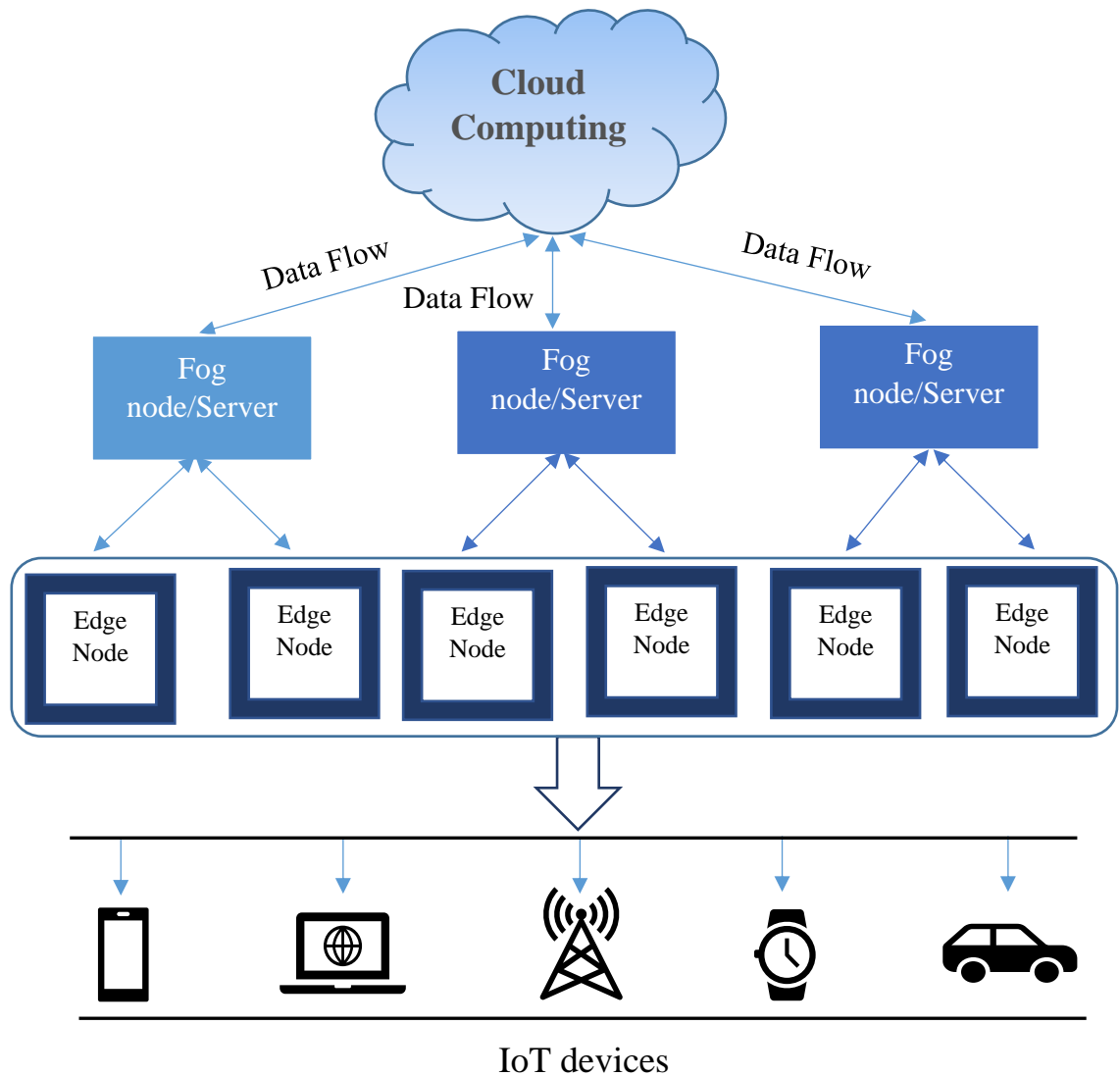

**Fig. 2.** IoT with Cloud, Edge and Fog computing architectures.

## 4.1. Cloud computing

Cloud computing provides computer services, such as servers, networks, software, databases, and data analytics through the internet, with faster deployment and more flexible tools and supplies [75]. It is a software system that is mainly application-based and stores data on various remote servers that users can access via the internet from any part of the world. Services provided by these servers are controlled by third-party organizations. Cloud computing works by adapting various service models and deployment methods to suit the customer and their needs. For deployment methods, the cloud has private, community, public, and hybrid models [76], which are targeted toward a private user, the general public, a single organization, and multiple organizations, respectively [77]. Therefore, the cloud computing infrastructure has all the elements that are beneficial to a larger population. Cloud computing provides on-demand services, a massive resource pool, mobility, scalability, and multitenancy all in a cost-effective solution, which makes it perfect for public and organizational use [77]. However, Alam [78] reported that in cloud computing, there is a chance that crucial components might not be available in times of need due to regional and business regulations. Also, there is always a risk that the concept of sharing resources might compromise the security and integrity of the information and contribute to a lack of confidentiality. Nevertheless, cloud computing has already shown its advantages in the fields of E-learning, E-governance, research, data storing, and so on [77].

As cloud servers in cloud computing are typically untrusted, it has become necessary to ensure the confidentiality of information and classifiers. To solve this issue, Li et al. [79] designed a paradigm for privacy-preserving outsourced classification. This paradigm includes a proxy homomorphic encryption mechanism that was developed on Gentry's scheme for the protection of sensitive data. In this mechanism, numerous data suppliers outsource completely homomorphic ciphertexts (encrypted data) to the evaluator ("S"), which stores and processes those ciphers. The evaluator "S" and the crypto service provider were combined to create a classification algorithm on data encrypted using distinct public keys. This model was also encrypted and stored in evaluator S, which can be utilized to serve clients with a prediction platform. The proposed algorithm proved to be semantically efficient in the encryption and prediction of data, which is also secure. However, the scheme fails to show the interaction between the crypto service provider (CSP) and evaluator S, and the cost of the communication is also not very favorable.

In cloud computing, the variety of service providers makes it a major challenge for enterprises to opt for a suitable cloud service that may meet their needs. As a solution, Abdel-Basset et al. [80] proposed a neutrosophic multi-criteria decision analysis (NMCDA) method based on the analytic hierarchy process (AHP) for evaluating the quality of various cloud services that can help a client to estimate distinct cloud services. The reason behind choosing a multi-criteria decision analysis is that various factors need to be considered while evaluating a service provider. In the prior study, the researchers had a board of expert decision-makers and also improvised the consistency degree of the metric by modelling an induced bias matrix in a neutrosophic setting. The cloud service estimation method was characterized by triangular neutrosophic numbers. In comparison matrices, various linguistic variables were used to represent those numbers. As an outcome, the proposed neutrosophic multi-criteria decision analysis method has resulted in numerous benefits for dealing with unclear and inconsistent data, while many organizations have confirmed the good applicability of this method. However, as a new approach, the amount of involvement from companies is very low, so the result of a broader assessment is unknown.

## 4.2. Edge computing

Edge computing is an integrated platform that combines network, processing, storage, and application key competencies at the edge of a network that is physically near the source of the data [81]. The point at which the edge analysis takes place is referred to as an edge node. The node can be anything between the data generating source and the center of the cloud that has processing and network capabilities [82]. Rimal et al. [83] used mobile phones and gateway as examples of edge nodes, with the first node connecting a person to the cloud center and the second connecting a smart home to the cloud center. Edge computing works by putting the data through three of its layers, namely the ending, cloud, and edge layers. The very first layer, which is the cloud layer, is in charge of scheduling both the nodes and cloud computing centers. It follows a control policy for scheduling to ensure better client service. In comparison to other paradigms, this also guarantees that data and computing are shared inside the network during decision making, rather than being sent to the central server [82]. Hsu et al. [84] reported that the edge layer is where all the nodes are used, and is designed to expand cloud services to the very edge of the network by connecting to both the cloud and the end layer. It is in charge of transferring information to a cloud layer continuously. To fulfil the low latency along with large traffic demand, this layer also provides three key functions: i) data caching, ii) localization computing, and iii) wireless access [85]. The end layer is the last layer that is the closest to end-users and includes devices. It takes the data that is put through the devices and sends it to the other layers for processing [86].

Edge computing works at the end of the network and within the network. Distributing work in edge nodes and cloud centers, edge computing minimizes the data load on the cloud center and makes it safer. It is more secure because the level and volume of data at risk can be reduced as the majority of data is processed on edge nodes rather than cloud centers. Therefore, if data are damaged at the end devices, it does not affect data on the center. It also lessens the burden of data on cloud centers by locally processing data and improving the overall data flow [82]. Although there are not any notable disadvantages to edge computing, Shi et al. [87] discussed the lack of edge computing used in service management and suggested some of the improvements cloud computing needs in service management, data abstraction, and security of the users. As an example, a standardized naming scheme for applications with a unique structure and service method is necessary for edge computing for proper communication, programming, addressing, object identification, and data transmission. Another development that should be considered is programmability in edge computing because the heterogeneous edge nodes in the network make it hard for users to deploy an application in the paradigm.

Nowadays, edge computing is used in mobile and data safety, attack detection, privacy preservation, vehicle, transportation safety, resource management in various paradigms, and so on [88]. It is also applied in real-time and context-aware situations, such as emergency healthcare and service recommendations. As the distribution and storage of information across various cloud devices are key concerns in today's age of massive amounts of data, edge computing is not immune to these issues. To address the latency-aware distribution of data copies at the edge of any network, Aral and Ovatman [89] proposed a distributed information/data dissemination strategy named D-ReP. It is based on the dynamic creation, replacement, and removal of the data replicas led by the continual analysis of data requests received from the nodes of the edge network. The procedure is divided into two versions, namely the source and edge, which iteratively moves replicas from the central storage towards the requesting edge nodes. In close proximity, if a cluster of edge nodes repeatedly requests the same pieces of data, replicas of the data are formed and relocated towards those nodes. The D-ReP model proposed by Aral and Ovatman [89] combines both versions. Specifically, the source version operates only in the node where the central storage is located and can only produce copies of data items in that adjacent and close node. The edge version operates at each node that has at least one copy and can be turned off when there are no replicas remaining. It also provide a messaging system to locate the replicas. According to the results, D-ReP can achieve 26% more delay reduction with 14% less incremental expense compared to the non-replicated sources of data with client-side caching. Aside from that, communication inefficiency and misunderstanding errors induced by replica deployment and detection were also shown to be insignificant. However, the proposed scheme lacks assurance in real-time performance, which might not be as efficient when used in emergency healthcare systems or self-driving cars.

In edge computing, the transport of massive data at the edge causes latency, which is in opposition to the immediacy demanded by many ubiquitous applications. To avoid that, Breitbach et al. [90] presented a data management technique for edge computing settings that separates the task of data placement from "task scheduling." The scheme provides a multi-level scheduler that assigns data to the system's resource suppliers while taking various contextual factors into account. The scheduler assigns tasks depending on the current context and also monitors the state throughout the runtime. The system modifies the amount of data copies needed to improve the trade-off between the latency in execution and extra expenditure for data management. The context-aware multi-level scheduler works by incorporating 4 data placements, 3 task scheduling, and 3 runtime adaption methods. After using the scheduler on a real-world test-bed, a prototype model tailored for the Tasklet model was tested to assess the efficacy of this model. The results demonstrated that a context-cognizant replication technique, combined with task scheduling based on performance awareness and dynamically changing runtime adaptation, outperforms all other existing models. The combined technique produced a task response time similar to that of complete replication while needing fewer data overhead. However, as promising as this approach appears, it has not yet been applied on a large scale to assess its full efficacy.

### 4.3. Fog computing

The fog computing architecture provides capabilities of computing, processing, storing, and networking services that are assigned across multiple end devices. As a result, it might be described as the opposite of traditional cloud computing [91]. It works as a bridge that brings the cloud and the end devices closer. This is accomplished by bringing computing, storage, and networking resources in close range to the end devices [92]. Therefore, fog computing extends the work of cloud computing by carrying out some of the processing near the very edge of the network, which in this case is the end-users. The key feature of this paradigm is its topology, consisting of geographically spread nodes that conduct the computational work and provide the networking and storing functions [92]. The devices added to the network are known as fog nodes, which can be placed everywhere if there is a network connection. To be a fog node, a device should be able to perform three core required functions: i) computing, ii) storing, and iii) connectivity [93]. Fig. 3 shows the difference in the combination of cloud and edge computing architectures with or without fog computing architecture. Therefore, the IoT solution, which refers to a fully integrated bundle of technologies attempting to solve a problem or generate a new organizational value, necessitates the usage of all three architectures.

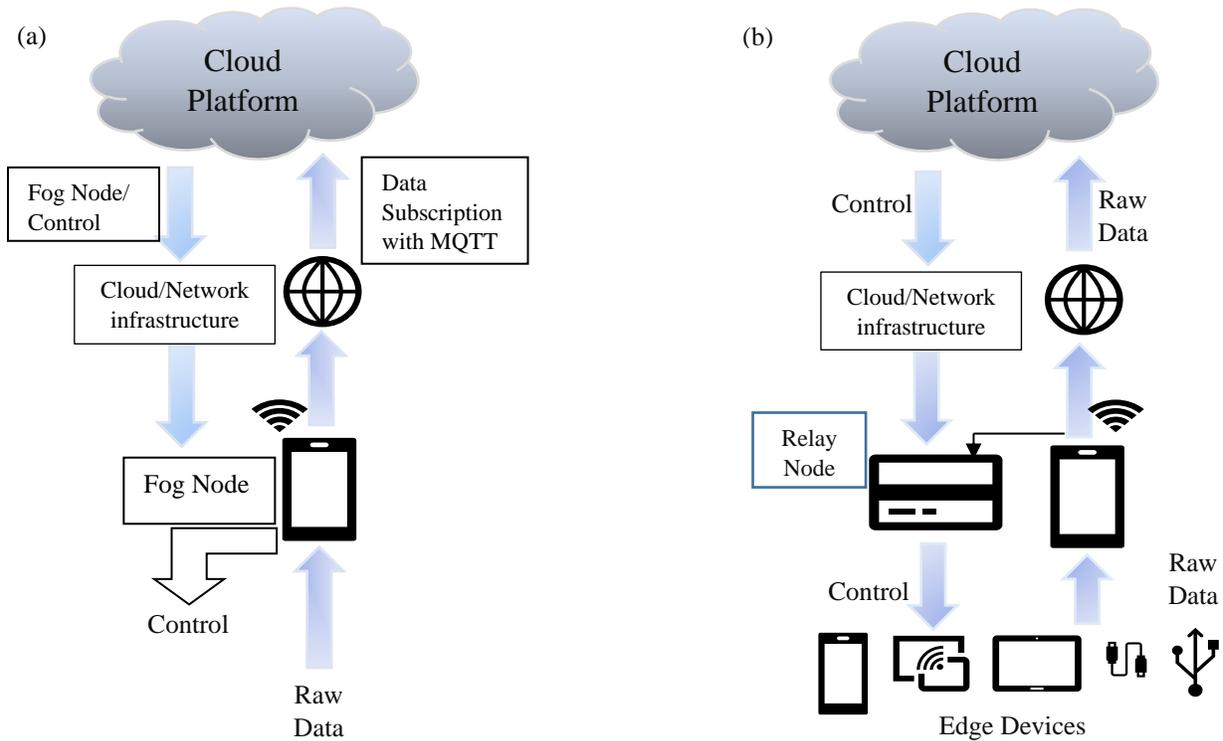

**Fig. 3.** Combination of cloud-edge computing architecture (a) with and (b) without fog computing architecture.

Since fog computing works as an intermediary system, it can enhance bandwidth and privacy by processing data or answering data, reducing the volume of data that is to be sent to the data center, and analyzing various sensitive data locally [93]. It can also reduce latency by working close to end-users and can build scopes for scalability. However, it has limitations due to its proximity to the edge of the network. While working as the intermediary, it may face intruder attacks, authentication issues, distributed denial of service, and so on, [94]. Additionally, the work distribution in nodes from a central point also starts access-control and resource management issues [21]. Fog computing is now being used in healthcare systems, service improvement, web optimizations, road safety, video processing for surveillance purposes, resource management in micro-data centers, data processing, and many other applications [95]. Fog computing often faces delays in execution time because there is no efficient scheduling and resource distribution of users' tasks. To efficiently schedule tasks and diminish this issue of delay, Rafique et al. [96] proposed a unique hybrid model of resource management called an innovative bio-inspired hybrid algorithm to offer effective consumption of resources in the fog-IoT paradigm. The suggested approach includes modified particle swarm optimization (MPSO), which distributes the load among the fog nodes, and the modified cat swarm optimization (MCSO) parts, which maintain all the usable fog resources. All simulations were performed with iFogSim, and the loads were allotted and managed based on the volume of user traffic. In contrast to the usual scheduling strategies proposed for fog computing, the results demonstrated that the suggested strategy offers promising outcomes for reducing energy usage, completion time, average reaction time, and cost. However, when applied in a fog-IoT environment, the learning strategies of this suggested model need to be reframed and reinforced.

Due to the fact that users' data is completely stored on cloud servers in fog computing paradigms, they lose authority over their data and face privacy issues. Additionally, traditional privacy and confidentiality protection solutions are incapable of preventing an attack within a cloud server. Wang et al. [97] proposed a three-tier storage framework that can fully utilize cloud storage while protecting data privacy. In this scheme, data are saved in the cloud server, fog server, and local device of a user. The data on the user's device is encoded using the Hash-Solomon code technique, which preserves the fragmented data, unlike other coding techniques. It also works to partition data into distinct sections for better use of storage. Apart from that, using computational intelligence, this program can estimate the distribution proportion that is stored on the fog, cloud, and local devices [55]. The Cauchy matrix was found to be the most efficient for coding among other compared matrices. Based on the test results, this approach can effectively complete the encoding and decoding parts without affecting the efficiency of cloud storage and can be assumed very effective in the case of maintaining security. However, the efficiency of the coding matrix with enormous data and real-time situations has not yet been analyzed, which may be impossible due to the fact that encoding massive data can affect the storage efficiency and cause latency. Table 3 summarizes the proposed algorithms for enhancing fog, cloud and edge computing architectures.

**Table 3.** A comparison of the proposed enhancement models for cloud, fog, and edge computing architectures.

| Computing Architectures | Suggested scheme/ Algorithm | Study | Main task(s) | Strength(s) | Weakness(es) |
|---|---|---|---|---|---|
| Cloud computing (CC) | Privacy-preserving outsourced classification | [79] | Protecting the confidentiality of data and classifiers in CC | Efficient data encryption and prediction | High communication cost, unknown interaction between CSP and evaluator |
| | Neutrosophic multi-criteria decision analysis | [80] | Helping clients select suitable cloud services | Good applicability found in real-world situations | Few companies have been considered in the evaluation of applicability |
| | Hybrid Genetic-Particle Swarm Optimization (HGPSO) | [98] | Scheduling tasks based on priority for improved utility and task execution time | Considerably better execution time, scalability, and availability | The cost of implementation has not been analyzed. |
| | Multi-objective Antlion Optimizer | [99] | Task scheduling focuses on multiple objectives, such as makespan minimization and utility maximization | Efficient in scheduling problems of large scale | No performance comparison has been done with recently proposed models for similar issues. |
| Fog computing | NBIHA - hybrid of MPSO and MCSO | [96] | Schedule tasks and resources to diminish delay | Lower cost, energy usage, completion time, | Not ideal for IoT environments |

|  |  |  |  | and average reaction time |  |
|---|---|---|---|---|---|
|  | Three-tier storage framework | [97] | Maintaining data privacy while taking full advantage of cloud storage | Good security and cloud storage efficiency | The possibility of latency while working with bulk data has not been analyzed. |
|  | A decentralized algorithm | [100] | Computational tasks allocation among the devices for efficient resource management | Remarkable system performance with a low signaling overhead | The algorithm was tested fully on a game model. |
|  | Hybrid IWO (invasive weed optimization) - Cultural Evolution Algorithm (CEA) | [101] | Energy-conscious work scheduling to reduce energy use and completion time | Improvement of energy consumption in some existing algorithms | No overhead issues were taken into consideration. |
| Edge computing | D-ReP: Distributed information/data dissemination strategy | [89] | Latency-aware distribution and storing of information | A good amount of delay reduction with less cost | Lacks the assurance of real-time performance |
|  | Multi-level scheduler | [90] | Avoid latency in transporting a large amount of data | good task response time with less data overhead | Needs assessment in large-scale application |

| | | | | |
|---|---|---|---|---|
| Prediction-mapping-optimization heuristic algorithm | [102] | Resource-requirement-aware server placement for better server selection | Cross-regional resource optimization scheme for cost minimization was also provided. | The accuracy of prediction cannot be guaranteed unless tested in the real network. |
| Deep Deterministic Policy Gradient (DDPG)-based algorithm | [103] | Solving computation offloading issues to reduce delays in heterogenous edge computing servers | 50% reduction of task delay and within deadline task completion | The task distribution is not cost-conscious. |

## 5. IoT integration with cloud, fog, and edge computing

### 5.1. IoT integration with cloud computing

IoT and cloud computing can be considered for rapidly evolving services with distinct properties. The IoT is based on smart gadgets connected with one another through a worldwide network and flexible infrastructure, which also enables scenarios involving ubiquitous computing. Hany et al. [104] discussed the differences between IoT and cloud computing and how integrating IoT with cloud computing can solve the majority of IoT concerns. IoT is characterized by widely dispersed devices with low processing and storage capacity. These devices can also confront issues related to privacy, performance, and security [105]. Generally, IoT can take advantage of the cloud's virtually unlimited assets in order to solve its technological limitations, such as data processing, communication, and storage. Thus, the cloud can be an excellent solution for managing and composing IoT services and developing applications as well as services that utilize the produced data [106]. The cloud can be benefited from IoT by expanding its chances to cope with actual things in a more flexible and distributed manner [107]. In many circumstances, the cloud can serve as an intermediary layer between models and applications, simplifying implementation and allowing for more robust functionality. Botta et al. [107] conducted a study by integrating two technologies and proposed the Cloud-IoT paradigm, which basically addresses the possible issues regarding its applications. The study highlights the vision of integrating both IoT and cloud computing as well as the challenges and availability of platforms to implement this paradigm. The authors analyzed in greater detail the issues related to this integration. Despite the comprehensive evaluation, the study did not investigate or include the open-platform issues.

Another study by Díaz et al. [108] on cloud computing and IoT integration focused on three unique elements, such as cloud infrastructure, cloud platforms, and IoT middleware. Cloud of Things, a combined paradigm of IoT and cloud computing, could address and resolve constraints related to data analysis, accessibility, and computation. Different existing levels of integration were illustrated with an emphasis on embedded IoT devices, e.g., to exchange information, and technologies used by IoT, such as RFID and WSAN. By incorporating cloud computing, the limitations of these IoT-related devices can be alleviated. Stregiou et al. [22] conducted a survey based on the most common issues related to the security of IoT and cloud computing integration systems. The benefits via the enhanced IoT capability by integrating cloud computing were also demonstrated in this study. An algorithm model consisting of two encryption algorithms, including

Rivest-Shamir-Adleman (RSA) and Advanced Encryption Standard (AES), was proposed in order to investigate the security concerns and challenges of the IoT and cloud computing integration. However, the authors did not demonstrate if the employed algorithms used to construct the security model can be employed to circumvent complex security challenges in the future. Cavalcante et al. [109] overviewed the development of IoT integration with cloud computing and suggested a direction for future research. The authors highlighted four topics: framework, platforms, architecture, and middleware. Although the issues associated with the standardization of integrated IoT and cloud computing services as well as handling actual and complex data were illustrated in the study, other challenges were not addressed.

## 5.2. IoT integration with fog computing

Integrating fog computing with IoT will benefit a wide range of IoT applications, as it allows for real-time communication among IoT devices. As such, latency is reduced, which is especially important for time-sensitive IoT applications [110]. Fog computing can enable large-scale sensor networks, a major issue posed by the proliferation of IoT devices, and thus assist a large range of IoT applications. [111]. Chiang and Zhang [112] provided an overview of IoT integration with fog computing by addressing emerging challenges in the systems of IoT and how the current system models are incapable of resolving the issues without fog computing. Fog computing integrated with IoT presents various advantages and can solve problems regarding controlling automotive applications, flight financial trading, and implementing end-to-end latency of around milliseconds between the control mode and sensor. It can also help to manage networks, systems, and applications of end-users. Aside from that, the study demonstrated how fog-based IoT may assure network scalability and improve radio access network (RAN) performance. However, the authors did not account for the difficulty posed by the constant failure of fog-based devices because of the fog computing's decentralized nature, which may result in the cessation of user activity, software, and hardware.

Puliafito et al. [113] illustrated the mobility support issue in the fog-based IoT system, concentrating on mobile IoT devices utilizing fog computing, and explored potential obstacles. In addition, three unique scenarios based on the IoT integration with fog computing were presented, and the need for mobility support was emphasized. In order to overcome the mobility challenges, various research directions were suggested for the future, such as proactive and reactive service migration, proper selection of virtualization, and migration strategies that might lead to enhanced performance and implementation of mobility support. The integration of 5G mobile networks was also explored and showed to be a viable research avenue to highlight fog-based mobility support systems. Nevertheless, the authors failed to address the mobilization and system management challenges. In order to identify security-related issues in the existing fog computing models, Khan et al. [95] analyzed the applications of IoT-based fog computing. The work demonstrated that the majority of the applications prioritize functionality over security, despite the fact that security is a fundamental component of the system. As a result, various fog-based platforms become vulnerable. The study mainly focused on assessing the impact and importance of security challenges and potential solutions capable of providing future security-based approaches for fog computing. However, system-level difficulties associated with fog computing, including service-oriented computing and resource management, were overlooked in the study.

## 5.3. IoT integration with edge computing

A massive transformation can be seen in both IoT and edge computing systems. Edge computing can notably assist IoT in resolving several complex issues, thereby enhancing its performance. As a result, it has become quite evident that both IoT and edge computing should be integrated. IoT, in particular, necessitates rapid responses over high and complex computational demands and vast storage space. To meet the needs of IoT applications, edge computing delivers an appropriate and suitable computational capacity, sufficient space for storage, and rapid response time [114]. IoT can help edge computing by increasing the structure of the edge computing framework to comply with the flexibility of edge computing nodes [87]. Edge nodes can also provide services using devices with residual computing capability or IoT-based devices. Several studies have attempted to use cloud computing to support IoT; however, in many circumstances, edge computing surpasses cloud computing [115]. As the amount of IoT devices increases, the further integration of IoT and edge computing is anticipated. Gopika et al. [116] conducted an experiment to evaluate edge computing and its associated mobile gaming technologies for a specific use case. In this work, a resource-intensive three-dimensional application was used as a paradigmatic example, and the response delay in various deployment scenarios was analyzed. Experiments demonstrated the importance of edge computing in meeting the latency requirements of augmented and virtual reality systems. However, the study failed to address the issues of edge computing system integration. Since edge computing usually incorporates a combination of numerous platforms, servers, and network topologies, it is considered a heterogeneous system [107]. As such, it becomes complicated to program and carries out resource management and data transmission in diverse applications that run on heterogeneous platforms [117].

Edge computing has the ability to overcome the latency/delay issues of IoT systems. The two major components of any application are computing latency, which often refers to the overall time required to process any data, and computing capability [118]. The transmission of data between servers and embedded devices leads to transmission latency [119]. Over the past decade, numerous studies have been conducted to obtain a solution for minimizing the latency of IoT systems by combining them with edge computing. For instance, Rehman et al. [120] demonstrated a profound scheme for computational offloading in the system of mobile edge computing. With the help of this scheme, an appropriate virtual machine was found on mobile devices for the rapid completion of tasks. As a result, the execution time, as well as power consumption, can be effectively reduced, resulting in energy savings. Therefore, computational offloading can lead to the reduction of transmission delay with the help of edge networks [121]. Although this scheme can ensure prominent results in the recognition of several online activities, further research and investigation are needed for the applicability of the proposed framework.

Since IoT uses an increasing number of sensors, the amount of data generated is enormous. It is almost impossible for the data to get transmitted to cloud servers directly without being compressed or processed. In order to avoid issues with transmission delays, this massive amount of data necessitates a maximum network bandwidth [122]. Before transmitting data to remote cloud servers, IoT gateways must therefore perform data pre-processing and aggregation. The objective is to manage and control traffic flow by automating data processing as efficiently as possible to lower end-user bandwidth requirements while preserving data quality [123]. Numerous studies have focused on the bandwidth of IoT. By deploying edge computing, Papageorgiou et al. [124] suggested a framework extension for stream processing that takes into account interactions with users, databases, and more entities, also known as topology-external interactions. This architecture allows for reduced bandwidth use and eliminates latency violations. However, further research is needed to reduce the bandwidth consumption of cloud-to-edge computing.

The surveyed literature demonstrated that future quality of life, cultural diversity, and human-device interactions will all be affected by the IoT technological revolution. New business models and opportunities will emerge as IoT is integrated with cloud, fog, and edge computing. However, the services and devices of the IoT will not work as expected or may even cause issues for users if the underlying network infrastructure is not effectively designed. Companies of all sizes have started allocating resources to IoT projects. IoT integration has the potential to boost a system's efficiency, effectiveness, and communication. For instance, the limitations of data analysis, accessibility, and computation could be addressed and alleviated with the help of the Cloud of Things, a combined paradigm of IoT and cloud computing. Edge computing can outperform cloud computing in many scenarios, and as the number of IoT devices grows, so does the level of integration between edge computing and the IoT. Edge computing provides sufficient computational power, storage space, and response time to suit the needs of IoT applications.

Fog computing with IoT can solve some of the problems caused by cloud computing, such as the management of time-sensitive IoT applications like gaming, simulation, and streaming. By integrating fog computing with the edge, data transfers and communication delays to the cloud can be minimized. Benefits and problems related to managing automotive applications, financial trading in aviation, and implementing end-to-end latency of approximately milliseconds between the control mode and sensor can also be addressed by combining fog computing with IoT. Fog computing integrates many different types of cloud computing hardware into the IoT. It has tremendous data processing capabilities due to its usage of massive amounts of storage and its ability to bring cloud computing to the network's edge. As the number of devices connected to the IoT continues to grow by the billions, fog computing has proven capable of keeping up. Although fog computing can quickly adapt to new circumstances, it can be difficult to adapt the IoT's workflow in an ever-evolving manner. Table 4 provides an overview of IoT integration with cloud, fog, and edge computing.

**Table 4.** Overview of the studies conducted on IoT integration with fog, cloud, and edge computing.

| Study | Main Task(s) | Advantage(s) | Disadvantage(s) | Application |
|---|---|---|---|---|
| Botta et al. [107] | Propose a cloud-IoT paradigm | Comprehensive analysis of open platforms and applications | Lack of open service platforms in multi-network | Healthcare, traffic management |
| Díaz et al. [108] | Analysis of current cloud integration frameworks | Overview of different existing levels of integration | Does not consider recent cloud-based platforms | Data backup and sharing, resource service |
| Stregiou et al. [22] | Investigate the security of the IoT and cloud computing integration system | Two encryption algorithms to survey the security issues | Cannot deal with the complex issue | Navigating security issues and threats, protecting cloud datasets |
| Cavalcante et al. [109] | Overview of cloud-based IoT integration development | Highlights integration challenges as well as providing future research directions; handles complex data | Does not highlight other complex issues about cloud-based IoT platforms | Cloud-based architecture and middleware |
| Chiang and Zhang [112] | Overview of opportunities and challenges of integration of IoT and fog computing | Ensure scalability of peer-to-peer networks and the performance of RANs; control applications; implement end-to-end latency | Does not consider fog-based breakdown challenges | Vehicle management, financial flight, trading |
| Puliafito et al. [113] | Mobility support in the fog-based IoT system | Overcomes mobility challenges and provides future research directions | Does not take into account mobilization and system management issues | Mobile cellular network (5G), call delivery service |

| Khan et al. [95] | Security issues in fog-based IoT models | Brings attention to the need of combating security threats; solutions of security-based directions of fog computing | Does not consider other fog computing system levels | Fog-based security models |
|---|---|---|---|---|
| Gopika et al. [116] | Resource-intensive three-dimensional application as a paradigmatic instance for response delay evaluation | Edge computing fulfils latency requirements. | Does not highlight issues of edge computing integration system | Edge-based mobile gaming technology |
| Rehman et al. [120] | Computational offloading scheme in the mobile edge computing system | Reduction of transmission delay by edge-based computational offloading; energy savings | Skeptical about the applicability of the framework | Mobile edge computing system |
| Papageorgiou et al. [124] | Edge-based framework extension for stream processing | Bandwidth consumption can be decreased. Violations about latency can be abolished. | Insufficient research for bandwidth consumption elimination | Manage and control traffic flow |

## 6. Applications of IoT computing integration

With the advent of smart devices and their subsequent improvements, the prospect of linking ordinary tasks over data infrastructure has gained a lot of traction. IoT has been boosted by improved technologies in WSN, ubiquitous computing (UC), and machine-to-machine (M2M) interfaces [125,126]. Automated systems must be integrated efficiently from a safe, technical, and economic aspect. Sensor devices, the internet of everything with trustworthy electrical installations, and comfortable environments will achieve higher productivity, cost depletion, health and safety for companies, institutions, and people [127]. Because of the widespread contemplation of profuse interest groups, remarkable applications of IoT have been introduced in various spheres: video surveillance, smart homes, smart metering, smart cities, smart healthcare, and environmental monitoring.

### 6.1. Video surveillance

The process of monitoring a site and searching for improper actions is known as video surveillance [128]. Machines adept at simultaneous recording and analyzing video data are already available, and they can equip security administrators with valuable insights rather than discrete bits of data. Video surveillance is commonly used for monitoring operations, remote video monitoring, facility protection, vandalism deterrence, employee safety, loss prevention, parking lots, public safety, outdoor perimeter security, traffic monitoring, and observing public sports events [128,129]. Rego et al. [130] introduced an artificial intelligence method for identifying and fixing faults in multimedia transference in a surveillance IoT network integrated through an SDN. Results showed that jitter can be minimized by up to 70%, and losses can be lowered from approximately 9.07% to 1.16%. Moreover, it exhibited 77% accuracy in detecting essential traffic. The system efficiency can be enhanced by using end-user interactivity or a broader data set. Gulve et al. [131] proposed a system that alerts users about the appearance of an individual on the surveilled premise while ensuring safety by recording that person's activities. The system's procedure begins with motion detection, which is refined to human detection, followed by alerting a number of nearby people or neighbors. According to the findings, the motion detection speed ranges from 3.1 to 9 seconds, and human detection takes 2 to 5 seconds with adequate accuracy. Besides security, the system is efficient enough to reduce power expenditure when the user forgets to shut down any electronic devices. Nonetheless, some more elements can be appended to the framework (i.e., face recognition) to switch the system on and off via a mobile phone.

Because IoT devices have constrained memory, power, processing capabilities, and resources, IoT networks are implemented in resource-restricted environments where edge computing is unavailable for security and data analytics. Stergiou et al. [132] suggested a stable approach for video surveillance systems functioning in resource-confined IoT settings to alleviate insufficient resources and data security issues. The four main stages of the proposed model are as follows: 1) saves only the video frames with detected activities, 2) encrypts the stored frames for reliable transmission, 3) synchronizes the encoded frames among sensor and cloud node, 4) eliminates the mediated frames from the sensor, and decodes the saved information on the cloud. During the experiment, the execution of resource-limited devices is evaluated on several types of sensor nodes. The suggested approach automates, accelerates real-time video data extraction speed, and inhabits less storage on the cloud in contrast to traditional strategies. Additionally, encrypting video frames maintains data reliability of video data while transferring from the sensor to the cloud. However, the production of ciphertext would be oversized for continuous video streaming in terms of the memory footprint that may be unviable to store and transfer data. Future security surveillance will effectively integrate three technologies that will entirely revolutionize the game, namely computer vision, automation, and deep learning, and are propelled by advanced hardware and IoT app cameras.

## 6.2. Healthcare

Today's society is entangled with numerous challenges affiliated with health issues, such as chronic diseases. The growing health concerns with higher healthcare expenses encourage the employment of computer-assisted technology for remote medical supervision [133,134]. Hence, extensive, convenient, and computer-aided technology is critical to meet the requirements for enduring healthcare and remote screening while also lowering financial and physical pressure on patients [135]. Kavitha and Ravikumar [136] proposed a four-module architecture consisting of a context-aware module (CA-M), IoT module (IoT-M), data pre-processing (DP-M), and decision-making module (DM-M), as well as some sensors and actuators. The findings show that the architecture produces superior outcomes to those individual modules in terms of accuracy, network latency, scalability, and reaction speed. However, the potential security vulnerabilities were not addressed in this study. Ray et al. [137] presented a cost-effective and wearable galvanic skin response (GSR) system for detecting the degree level of human physiological activity in smart e-healthcare systems by capturing, intensifying, and processing GSR data. The system was produced on a feasible Pyralux substrate, making it lightweight, portable, and flexible. The obtained data can be seen with moderate power consumption in smartphones. Nevertheless, privacy and security demands were neglected in the procedure. Furthermore, the smart Saskatchewan healthcare system by Onasanya et al. [138] was created for cancer care, business analytics and cloud services, operational services, and emergency service. It was built based on the technologies of IoT, particularly WSN and other linked devices, for durable care and uses a full-mesh hierarchical network topology to provide satisfying security while consuming little power. It also includes IoT-based designs, network mapping, techniques, platforms, applications, and architectures for IoT-based solutions for all linked things aimed at delivering efficient healthcare. Although the recommended IoT healthcare system has a high proclivity for optimizing operational processes value to ameliorate healthcare delivery, it poses significant operational risks, particularly in sensitive data and combined information with a range of smart linked devices.

Awaisi et al. [139] developed an identity management-based user authentication technique to improve system security and mitigate data breaches by dubious users. They used elliptic curve encryption for authentication and devised a machine partitioning paradigm in fog nodes to provide an optimal fog infrastructure for healthcare. It exhibited a significant reduction in latency and network use, making it ideal for latency-sensitive healthcare. Moreover, the technique achieves authentication through an identity management technique, such as Elliptic Curve Cryptography and SHA-512, to avoid security issues, the most crucial shortcoming in IoT-relevant healthcare applications. Similarly, Tuli et al. [140] proposed HealthFog, a fog-based system in HIoT that uses deep learning to detect cardiac problems autonomously. Using FogBus, a fog-based architecture, the researchers verified their system in several circumstances to prove real-time forecasting, low latency, low power utilization, low bandwidth consumption, and excellent accuracy. However, the expense of putting this framework in place was overlooked.

## 6.3. Smart home and smart metering

'Domotics' is the term for home automation or smart building [141]. IoT-based smart home tech has revolutionized human existence by proffering connectivity to every human with a smart device, notwithstanding location and time [142,143]. Smart homes are autonomous buildings with sensors and inaugurated controls, such as ventilation, heating, air conditioning, lighting, security, and hardware systems. IoT maintains the network connectivity of "gateways" that handle systems with a user interface communicating with a mobile phone, tablet, or PC [144]. Khan et al. [145] designed an IoT Smart Home System (IoTSHS) that enables smart home remote control over a mobile phone, computer, and inferred remote. IoTSHS is constructed with a WiFi-based microcontroller and a temperature sensor. It must be affixed to the things superintended by the power distribution box with switches or relays. For those who cannot use their smartphone to manage their appliances, the developed system may also help them. The system can be connected to WiFi and operated via a web browser, independent of the operating system. IoTSHS might connect to the home router, allowing the consumer to control the appliances while keeping an interconnection without purchasing, downloading, or installing any application. Therefore, the designed system can benefit the entire community by rendering a newfangled remote control for the smart home. However, IoTSHS has a cost and might not be affordable for many. Malche and Maheshwary [146] explain how to use the Frugal Labs IoT Platform (FLIP) to create an IoT-based smart home. The study covers smart home functions, applications, FLIP structure, and smart home service implementation utilizing FLIP. Specifically, the proposed technology is substantially required, flexible, and secure in monitoring and regulating the home automation system. Home air quality can be tracked regularly by this system, and notifications about potential health hazards may be sent to the user. The system can detect every movement in the home and allows the user to control the window and door locks. The system includes smart appliances, smart lighting, and a smart air-conditioning system that provides improved energy and resource efficiency, which can be deployed in smart cities in the future by adding new functionalities.

Hoque and Davidson [147] presented a low-cost smart door sensor architecture that notifies the client of door actions in their home and office via an Android application. While the researchers utilized multiple programming languages (Java, Python, and C++) to implement various aspects of the sensor, interference with the 433 Hz RF band may induce complications. Many household appliances utilize RF signals, therefore there could be numerous RF receivers attempting to send signals to the Raspberry Pi at the same time, or it could be hogging signals that it was not supposed to. An interference evaluation with the RF applications can be done to suppress this issue in the future [148]. Information and communication technologies are evolving to meet the needs of all clients in terms of energy savings. Consumers anticipate accurate energy measurements, rapid data transfer, and superior customer service. A multi-technology smart grid system is the most cost-effective approach. The consumer can obtain details about energy consumption using their device's IP address. The client code is deployed to check the client's information, such as their location, web server connection, disconnection, and content. This potential solution provides accurate and trustworthy information about electrical energy management systems (EMSs) via IoT. However, more progress may be accomplished in online control and energy monitoring from any place.

Over the last few years, numerous research has been done in the domain of the smart home and context-aware IoT. Major difficulties associated with smart homes are interoperability, energy-aware consumption, context-aware middleware, privacy, and security awareness. The solutions for such challenges have some similar attributes, including: web services support, security mechanism, user interface support, adding devices, devices/systems' configuration and remote access control, device discovery, context acquisition, context reasoning, abstraction, aggregation, service management, reusability, context dissemination, scalability and fault tolerance, remote control of switching devices, controlling energy wastes, authentication, integrity, performance optimization, encryption techniques, machine learning, user authorization, personal information, and access management [58,149–152].

## 6.4. Smart cities and communities

Services and infrastructure are necessary to meet the needs of urban residents due to the rapid population growth. Physical objects, such as any appliances, devices, cars, buildings, and other items integrated with sensing devices, software, and network access, can connect in the IoT with minimal or no human mediation to provide services to residents in smart cities [153,154]. In broad terms, a smart city is an urban environment that makes use of information and communication technology (ICT) and other associated smart technologies to maximize the productivity of conventional city quality of services (QoS) and activities given to city natives [155]. A smart city ensures the potent usage of resources and provides advanced services to inhabitants [156]. As the smart city is an IoT adaptation, it retains the IoT's core operating processes. IoT provides crucial building blocks for smart cities, such as data management, creation, and application handling [154,157]. The smart city includes smart homes, institutions, healthcare, parking lots, transportation, water, and weather management.

Several studies have introduced innovative models, frameworks, and technologies of IoT to make city and community life more feasible. For instance, to handle real-time transportation data, Silva et al. [157] created a model for evaluating transportation records using two open-source analytical engines, namely Hadoop and Spark. The model is composed of four layers, including network, data collection, acquisition, data processing, and application, whereby each layer is developed for processing and managing data in an orderly manner. Hadoop and Spark are used to test the data at the processing level. The data are distributed to a smart community by applying the endorsed mechanism. A variety of credible transportation datasets were used to assess the model's viability. Data processing and real-time transmission among citizens were done in the shortest time with high accuracy. This architecture is significant to implement in different vehicular network conditions in the future. Kanteti et al. [158] designed a smart parking system that allows the user to detect the nearest available parking spaces and lots in a particular area. By harnessing IoT's power and incorporating it with novel electronic computers and sensors, this concept increases the intelligence of the typical parking system. The smart parking platform reduces the time spent looking for parking lots and avoids gratuitous drives through congested parking lots, which is advantageous for trading city areas. The system can be executed using broad database storage, such as MySQL, Python, and cloud. A vehicle can be guided by applying the collected data via image detectors and sensors to the parking area, and the required directions are sent to the user's mobile phone. The recommended algorithm also offers more accommodation for vehicles, less power consumption, user-friendly parking, as well as solutions to numerous obstacles associated with vehicle parking. This algorithm is not only technologically efficient, but also gives a pleasant experience in terms of fuel economy, time-saving, and easy accessibility.

Concurrently, smart communities aim to ameliorate citizens' well-being and comfort in metropolitan areas utilizing IoT [154]. For instance, Ray and Goswami [159] advocated creating and executing smart water meters using cloud computing and IoT with machine learning algorithms to identify imprudent and standard water consumption levels in any location. The preservation of pure water supplies has become a global crisis in recent decades. Importantly, the recommended sensors in the cloud can facilitate hydraulic data monitoring, alarm notifications, and automatic control. A critical assessment of this research will enable one to take the appropriate decision. Thus, they recommended an intelligent water metering system that will aid in the reduction of water waste and help the community. As the researchers preferred a serverless architecture, the proposed meters can be conveniently adopted and integrated on a big scale. Ranjan et al. [160] offered an idea for smart rainwater collection by applying IoT. The model consists of two tanks that are separated by a segregation ratio of 60% to 40%. The rainfall detection sensor is mounted on the vertex of the system to identify if it is raining. The operating model was analyzed using two bottles of water (drinking water and acidic). The machine accurately measured the pH of the water and then stored it in separate chambers based on pH level. Moreover, the system could be monitored on a webpage through IoT. The inclusion of IoT offered the system an advantage over other conventional rainwater harvesting systems (RWHS). Additionally, IoT simplified this model by removing most of the obstacles of traditional RWHS. The United Nations Report forecasted that by 2050, urban areas will contain 66% of the world's population. As such, cities use a substantial proportion of the world's resources; in fact, cities utilize 75% of the total energy in the contemporary world [161]. As a result, sustainable, smart, and reliable cities must be constructed with IoT and sensor system applications. These "sensory cities" can inflate people's lives and management processes by focusing on environmental sustainability, energy conservation, and people's welfare as the three main pillars [162–164].

## 6.5. Environmental monitoring

 Environmental measurements via IoT provide pertinent information about the surroundings for productive…estimations, for instance, for remote management of cooling/heating equipment. This field relies on wireless sensor networks or remote sensing to acquire environmental data. Currently, IoT plays a vital role in environmental monitoring systems as well as in many other industries [165]. Velásquez et al. [166] developed three individual IoT-based wireless sensors to monitor the environment, including User Datagram Protocol (UDP)-based Wi-Fi, Bluetooth Low Energy (BLE), Wi-Fi and Hypertext Transfer Protocol (HTTP). Each sensor records data in remote areas, which can be viewed from any internet-enabled device, facilitating the inspection of broad geographic regions. The three proposed architectures can be implemented in monitoring applications considering their energy autonomy, solution complexity, flexibility, and internet connection facility and, thus, ideal for IoT solutions. The results reveal that Wi-Fi and BLE are the best monitoring technologies that can be used with the extensively used ZigBee protocol. However, BLE requires further protocol refinement and the addition of appending mesh network features to improve its performance.

Behera et al. [167] improved the Stable Election Protocol (SEP), which enacts a threshold-based Cluster Head (CH) selection. The threshold ensures that energy is evenly distributed between CH nodes and members. The sensor nodes are allocated into three classes (normal, intermediate, and advanced) based on their baseline power generation. Results indicate that the system exceeds distributed energy-efficient clustering (DEEC) and SEP protocols by 56% in throughput and by 300% in network lifetime. In the future, the updated method can be implemented in a mobile network in which nodes move at a steady rate from one point to another. Pasha [168] employed Arduino UNO, a Wi-Fi module to facilitate data processing and transmission to the Thingspeak cloud. The obtained parameters are saved in a cloud platform called 'Thing speak,' which allows the construction of a public-based channel to explore and evaluate the channel. The cloud exploits graphical visualization and makes itself available to users as a virtual server. Further, the objects and users interact with the cloud through potential 'wireless internet connections.' The IoT creates a connection and grants communication with all kinds of devices. The strength of this system is that the user can directly access and retrieve climatic parameters in JSON, XML, or CSV files. For direct access to the measured parameters, an Android application is also being developed.

Velásquez et al. [166] implemented an IoT-based approach for monitoring the environment by utilizing open hardware and open source tools that cost less than $150. This network analyzes temperature, carbon monoxide (CO), relative humidity, noise, particulate matter 2.5, and UV radiation with a reading frequency of every 40 s and hibernating duration of 15 min. After data are collected from the environment via sensors, they are stored in a MySQL database and published by a local web server Apache. Moreover, the system provides geotagged information and environmental degradation information. By allowing the user to modify nodes per their preferences, the system offers remarkable flexibility and scalability, allowing for better fiscal resource planning and strengthening community awareness of environmental issues. Table 5 provides a summary of the primary studies on IoT integration with computing platforms for various applications.

**Table 5.** Main surveyed studies conducted on IoT integration with different computing systems for various applications.

| Applications | Study | Architecture/module /sensors | Objective | Outcome/accuracy | Remarks |
|---|---|---|---|---|---|
| Video Surveillance | [130] | SDN, Classification method | Detect and resolve faults in multimedia transmission in an IoT surveillance context | It can reduce jitter by up to 70% and reduce losses from 9.07% to nearly 1.16%. Also, it is capable of recognizing critical traffic with 77% accuracy. | The system accuracy can be enhanced by using the end users' interaction or by using a more complete dataset. |
| | [131] | Raspberry Pi, Arduino, and GSM module | Implement IoT-based smart video surveillance | Motion detection speed: 3.1–9 s; Human detection speed: 2–5 s Required time to send SMSL: 20–30 s | Few more components can be added to the framework (i.e., face recognition) to activate and deactivate the system through mobile. |
| | [132] | Raspberry Pi, OpenCV, Advanced Encryption Standard (AES) | Provide a solution for a resource implementation framework that safely performs in constrained video surveillance circumstances | Works efficiently in a dynamic environment where activity levels fluctuate; useful for data extraction with lower processing requirements, and secure data transmission | The generated ciphertext in continuous streaming would be very large in terms of memory footprints that may not be viable to transmit and store. |

| Healthcare | [136] | IoT module (IoT-M), context-aware module (CA-M), data pre-processing module (DP-M), decision-making module (DM-M), BPNN and AGHO | Innovate an optimum neural network model for real-time health monitoring | Provides better outcomes with elevated scalability, accuracy, low response time, and network latency | Ignores the potential security threats |
|---|---|---|---|---|---|
| | [137] | Arduino Uno and Pyralux; Arduino IDE, makerplot, Fritzing; ATTiny85 and GSR sensor. | Assess and monitor IoT-guided human physiological galvanic Skin Response (GSR) level | It outperformed Shimmer3 over 7 different factors. | Does not address security and privacy requirements |
| | [138] | WSN, Full-mesh hierarchical network topology | Apply IoT and WSN based smart Saskatchewan healthcare | Can optimize operational processes' value to improve healthcare delivery | Has significant operational challenges in highly sensitive data |
| | [139] | Fog-based, elliptic curve encryption, SHA-512 | Leverage IoT and fog computing in the healthcare system | Decreases the latency and network usage value to a significant level and is best suited for latency-sensitive healthcare systems | Excludes the privacy breaches by using an identity management mechanism for user authentication |
| Smart Home and smart metering | [145] | IR, WiFi based | Develop a smart and intelligent parking system for city commercial areas | It provides advanced remote controlling for the smart home. | A significant number of families might not be able to afford the cost. |
| | [146] | Frugal Labs IoT Platform (FLIP) | Control smart homes based on FLIP architecture. | Monitors and controls the smart home environment efficiently | New functionality can be easily added to the system and can be implemented in smart cities in future. |
| | [147] | Arduino, Raspberry Pi 2, RF receiver-transmitter pair | Implement an IoT based security system for a smart home | The door sensor alerts the user about door open events through an Android application in the house or office. | Sensors could pick up unintended signals as many home devices use RF signals to communicate. |
| | [148] | Arduino Uno board and Ethernet | Implement smart metering based on IoT | The user can obtain information about energy consumption using IP addresses on their devices. | Online control and energy monitoring may provide more progress. |
| Smart cities and smart communities | [169] | Hadoop and Spark | Design a smart transportation system | Data processing and real-time data transmission happen within a short duration with high accuracy. | Can be applied in various vehicular scenarios in future |
| | [158] | sensors and image detectors | Design a smart parking system for smart cities' business areas | The algorithm provides increased vehicle space, user-friendly parking, and reduced energy use. | It gives a pleasant experience regarding fuel economy, time-saving, and less power consumption. |

| | [159] | Cloud computing, machine learning, IoT | Innovate a smart water meters | Can measure the standard and excessive use of water | Conveniently adopted and integrated on a big scale |
|---|---|---|---|---|---|
| | [160] | NodeMCU module WiFi. | Invent a smart rainwater collection harvesting system | The system stores water based on pH level and allows the users to monitor rainwater on a webpage via IoT. | The system has an advantage over other conventional rainwater harvesting systems since it utilizes IoT more than other |
| Environmental monitoring | [148,170] | UDP-based Wi-Fi; Bluetooth Low Energy (BLE); Wi-Fi and HTTP | Analyze IoT-based wireless sensors for environmental monitoring | WiFi and BLE are the most appropriate for monitoring. | The protocol refinement and the addition of mesh networking features to BLE devices will make them even more attractive and productive. |
| | [167] | Stable election protocol (SEP) | Design an improved routing protocol for environmental monitoring | Exceeds DEEC and SEP protocols by 56% in throughput and 300% in network lifetime | This method can be implemented in a mobile network, where nodes can move at a steady rate from one point to another. |
| | [168] | Arduino UNO | Create a Thingspeak-based monitoring and sensing system | Users can directly access climatic parameters and retrieve them in JSON, XML, or CSV files. | An Android application is also being created for direct access to environmental information. |

| [166] | Open hardware and open-source tools | Prepare a low-cost environmental monitoring system | Analyzes temperature, carbon monoxide (CO), relative humidity, noise, particulate matter 2.5, and UV radiation and publishes measurements on the local web server | Allows a user to modify nodes according to their preference, thus ensures remarkable flexibility and scalability at a low cost |
| --- | --- | --- | --- | --- |

## 7. Advantages and challenges of IoT integration with Cloud, Edge and Fog computing

The IoT and cloud computing systems are two quickly evolving technologies with unique features. IoT is composed of devices that have a dynamic infrastructure and connect with one another through a worldwide network system. Cloud computing is built with a massive network system that has a strong processing capacity and almost infinite storage [171]. As the IoT has some limitations like weak processing ability and limited storage, it must address issues, such as efficacy, safety, confidentiality, and dependability [107]. Therefore, the best way to handle these issues is to diffuse IoT with cloud computing. Since IoT is capable of working with real-world objects and scenarios in a dispersed manner, cloud computing can also benefit from IoT by connecting with a wide range of device systems [172]. The IoT is already connected with billions of devices and generates big data. Cloud computing has the ability to deal with such enormous amounts of data generated by IoT because of its unlimited storage. In addition, the IoT-infused cloud creates new possibilities to develop new services and products by interacting with IoT real-world scenarios [173]. However, apart from the benefits, there are a number of challenges to IoT integrated cloud services. For instance, the real-world data from IoT are sent to cloud computing via cloud-based IoT. Even so, the most pressing issue is that this method is unreliable as there is no method to handle sensitive information due to the lack of strict guidelines and regulations that threaten the privacy of users [174].

Edge computing has the advantage of being able to handle massive data, services, and computational applications, which can be moved from the central hub to the edge of the network. Utilizing existing resources, edge computing manages and stores data while also enabling control of various activities [175]. This feature of edge computing helps IoT to make the most out of pooled edge computing. As the processing of data and storage are performed closer to the end-users, it is easier to experiment with the data in a faster and less expensive way using an IoT integrated edge computing system and make better decisions quickly [87],[176]. Even though the proximity of the system and nodes to the end-users is one of the advantages of edge computing, it is also a disadvantage. Consequently, edge computing cannot be used remotely and has poorer computational abilities than cloud computing [177].

IoT integrated with cloud computing is more popular because of its big data storage and accessibility. However, it also presents a few challenges that can be addressed by fog computing integrated with IoT. For example, cloud computing has issues with dealing with IoT time-sensitive applications, such as video gaming, simulation, and streaming [178]. Fog computing connects IoT divisions with a wide range of cloud computing devices and can bring cloud computing to the edge of the network. Because it utilizes large amounts of storage, it has a strong data processing capability. Fog computing also connects the users' end devices to traditional cloud services through a virtualized paradigm of processing, memory, and networking devices [179]. Therefore, fog computing integrated with IoT can deal with time-sensitive applications, greatly benefiting IoT devices.

Fog computing minimizes the latency issue, which is essential for time-sensitive applications, by enabling instantaneous interaction between the IoT devices [180]. Since the IoT networking system grows by connecting with billions of devices, fog computing is able to keep up with the large-scale IoT networking architecture. Despite the potential of fog computing to dynamically adapt to change, it is difficult to continuously adjust the workflow structure of IoT. The internal quality and efficacy of IoT can also be affected if fog computing cannot support the dynamic change of IoT [181]. Furthermore, software and hardware deterioration affects portable devices, resulting in changes in workflow behavior and gadget attributes [12]. As a result, fog nodes' will require sophisticated and automated modification of their topological structure and resources. The fog networking architecture disperses nodes randomly at the edge, increasing the complexity of the network [182].

## 8. Future opportunities and directions

The IoT network, a fast-evolving speculative technology, can process a large amount of data in order to make intelligent decisions without human interference. The features of this network create a new generation of technology marked by automation and work in the development of artificially intelligent devices. The enabling technologies of IoT, assist us in the practical implementation of IoT systems and solutions prior to broad adoption. According to the researchers, more than 10 billion IoT devices were active in 2021 [183]. It is predicted that the applications will be expanded to cover smart grids and smart cities by 2025 [184] and that the number of active IoT devices will surpass 25.4 billion by 2030 [183]. IoT also helps to increase the effectiveness of data collection, which is why its integration assists organizations to operate efficiently by decreasing human errors [185]. With a range of supporting technologies based on cloud, fog, and edge architectures, the integration can be accomplished successfully. The integration of IoT with cloud, fog and edge computing is becoming increasingly prevalent in today's world. IoT is regarded as the most significant concept of the future internet, which is anticipated as a group of data communication network technologies that will be available in the future, bringing together the seamless networks and networked things into a single global IT platform [186]. Claiming both IoT and cloud computing as part of future networks, some scholars [187] report that cloud computing enables a backend solution for processing large data streams when everything will be connected through seamless networks. According to the 5G Observatory Quarterly Report of 2021 [188], 124 million IoT-supported 5G connections were added globally between Q1 and Q2 of 2021, indicating a 41% increase, which is due to the growth of IoT usage accessed through the cloud.

According to researchers, fog computing is at a critical stage of development, with the potential to reduce operational costs and handle some concerns of IoT, such as latency, storage, and data traffic. Fog computing architecture is thus required to provide a smart platform for managing the distributed and real-time properties of future IoT networks [189]. In addition, the edge computing architecture is also utilized in IoT integration with cloud and fog computing to solve issues in seamless networks. The edge computing architecture is anticipated to contribute to the future IoT by connecting a huge number of devices that generate massive amounts of data at a rapid rate, with applications demanding extremely low latency [187]. This architecture actually aims to enhance efficiency, minimize latency, consume less bandwidth, and improve security [190]. It focuses on the real-time applications of IoT that require a very immediate response. Therefore, the upcoming 6th generation networking system is predicted to be highly benefitted through the integration of IoT.

6G technology has the potential to revolutionize numerous industries and is thought to be the foundation for achieving the full potential of IoT. Examples of the integration of IoT in 6G include holographic teleportation, remote healthcare (telemedicine), smart cities and ecosystems, autonomous transportation systems, improved opportunities for distant learning, brain-computer interface, and other advanced technologies [191]. Table 6 presents the future of IoT and the opportunities in 6G networking systems. The expected exponential proliferation of smart devices, as well as the confluence of low-cost architecture, communication, and data, will propel IoT from a visionary concept to reality [192]. It is also true that the cybersecurity risks of IoT devices have not been adequately addressed over the last decades. The network can be put at risk as most IoT devices are not designed to fix security vulnerabilities (such as encryption failure, software defects, obsolete technologies, defective access control, and others) through regular updates [193]. To address these issues, the collaborators must use open standards to ensure that IoT is reliable, compatible, and can deliver secured services in a seamless manner [194]. Techniques that will minimize energy consumption even further must be implemented, such as green technology for IoT devices to make them more energy-efficient. By concentrating on solving such difficulties associated with IoT integration, better technologies in terms of speed, security and performance can be attained.

**Table 6.** Future of IoT and the opportunities in the 6G network.

| Sector | Future Opportunities | Future Direction |
|---|---|---|
| Industrial | – Better connectivity<br>– Reliable<br>– Timely communication of information<br>– Flexible manufacturing system<br>– Development of an intelligent Industrial IoT (IIoT) edge<br>– Working with persona-based IIoT | Increased network efficiency and the ability to coordinate, automate, deploy, and protect diverse use cases at hyper-scale will accelerate the industrial internet's development. |
| Healthcare | – Advancement: Improvement in patient health will be ensured.<br>– Better monitoring service<br>– Remote monitoring: Medical cards will be read in real-time by customized software.<br>– Assistance: Doctors will be assisted in doing a more detailed inspection of a patient's health.<br>– Wearables: The development of error-free devices will keep the patient's records updated correctly. | Further IoT improvements will result in even more low-cost treatment options for patients; also, having access to precise medical information gathering options may lead to fewer patient visits. |
| | – Asset monitoring: Connecting through the IoT devices reduces the chances of improper treatment.<br>– High data rates | |
| Transportation | – Telematics for ecology and economy<br>– GPS fleet monitoring<br>– Geofencing<br>– Real-time ubiquitous communication between the driver and the car<br>– Ultra-low latency | Increased usage of monitoring devices by legal authorities will aid in the reduction of various crimes, and the ability to use real-time data will ensure security. |
| Smart Grid | – Limited data logging rates<br>– Placement of Phasor Measurement Units (PMUs)<br>– Limited energy-load resolution<br>– Improved cybersecurity<br>– Limited data resolution<br>– Unlimited number of energy sensing nodes | Renewable energy costs will be decreased for all users as this technology increases production; for example, the cost of consuming fuel energy may decrease despite an increase in usage. |
| Smart City | – Public safety<br>– Improved smart building domain<br>– Cleaner and sustainable environment<br>– More applications of WSN technologies<br>– 15-minute city (urban regeneration model in post-pandemic cities)<br>– Extended Reality (XR)<br>– Smart water surveillance<br>– Unmanned aerial vehicle (UAV) techniques | Citizens can save time while also increasing their quality of life by connecting their smartphones or other smart devices to other technical things. |

## 9. Conclusions

IoT is designed to improve humans' well-being and the efficiency and intelligence of things. This study comprehensively surveyed the integration of IoT with cloud, fog, and edge computing architectures in depth to enable a variety of applications. In particular, the relationship between cloud, fog, and edge computing with IoT integration was investigated along with its future potential. Below are the major conclusions drawn from this survey:

- Cloud of Things, a combined paradigm of IoT and cloud computing, could address and alleviate data analysis, accessibility, and computation-related restrictions.

- By enabling instantaneous communication between IoT devices, fog computing reduces latency, which is critical for time-sensitive applications. It can also support large-scale sensor networks, a significant concern as the number of IoT devices increases.

- The integration of fog computing with IoT can result in a number of benefits and solve difficulties, such as managing automotive applications, aircraft financial trading, and implementing end-to-end latency of about milliseconds between the control mode and sensor.

- Fog computing connects IoT divisions to a wide variety of cloud computing devices. It can bring cloud computing to the network's edge, and because it employs enormous amounts of storage, it has great data processing capabilities.

- Since the IoT networking system is expanding by connecting billions of devices, fog computing is able to keep up with the IoT networking architecture's massive scale. Despite the capability of fog computing to react dynamically to change, it is challenging to continuously modify the IoT's workflow structure.

- To meet the requirements of IoT applications, edge computing delivers adequate computing capacity, sufficient storage space, and quick reaction time.

- The IoT technological revolution will have an impact on future quality of life, diverse lifestyles, and how we interact with humans, technology, and devices. Large corporations have already begun to invest in IoT initiatives. The success of IoT will depend on its architecture, as the services and devices will not function properly or may cause problems for consumers in the absence of a well-designed network infrastructure.

- Despite the wide range of concerns that need to be addressed, IoT appears to have numerous potentials. Integration of IoT requires additional research as IoT has the potential to significantly improve connectivity infrastructure as an inevitable component of the future internet.

## Acknowledgments

The authors highly express their gratitude to Asian University for Women, Chattogram, Bangladesh for their support in carrying out this study.

## Declarations

### Ethical approval

Not applicable.

### Competing interests

The authors declare that they have no competing interests, either financially or otherwise.

### Authors' contributions

Shams Forruque Ahmed: Conceptualization, Writing – original draft, Supervision; Shanjana Shuravi: Writing – original draft, Methodology; Shaila Afrin: Writing – original draft, Validation; Sabiha Jannat Rafa: Writing – original draft, Data curation; Mahfara Hoque: Writing – original draft, Formal analysis; Amir H. Gandomi: Reviewing and Editing, supervision.

### Funding

### Availability of data and materials

Not applicable.